\documentclass[pra,reprint,aps,superscriptaddress]{revtex4-1}
\usepackage[latin9]{inputenc}
\usepackage{amsmath}
\usepackage{graphicx}
\usepackage{natbib}

\makeatletter
\graphicspath{{./figures/}}

\DeclareMathOperator{\erf}{erf}

\makeatother

\begin{document}

\title{Atom-optics knife-edge: \\
 Measuring narrow momentum distributions}

\author{Ramón Ramos}

\affiliation{Centre for Quantum Information and Quantum Control and Institute
for Optical Sciences, Department of Physics, University of Toronto,
60 St. George Street, Toronto, Ontario M5S 1A7, Canada}

\author{David Spierings}

\affiliation{Centre for Quantum Information and Quantum Control and Institute
for Optical Sciences, Department of Physics, University of Toronto,
60 St. George Street, Toronto, Ontario M5S 1A7, Canada}

\author{Shreyas Potnis}

\affiliation{Centre for Quantum Information and Quantum Control and Institute
for Optical Sciences, Department of Physics, University of Toronto,
60 St. George Street, Toronto, Ontario M5S 1A7, Canada}

\author{Aephraim M. Steinberg}

\affiliation{Centre for Quantum Information and Quantum Control and Institute
for Optical Sciences, Department of Physics, University of Toronto,
60 St. George Street, Toronto, Ontario M5S 1A7, Canada}

\affiliation{Canadian Institute For Advanced Research, MaRS Centre, West Tower
661 University Ave., Toronto, Ontario M5G 1M1, Canada}

\date{\today}
\begin{abstract}
By employing the equivalent of a knife-edge measurement for matter-waves,
we are able to characterize ultra-low momentum widths. We measure
a momentum width corresponding to an effective temperature of 0.9
$\pm$ 0.2 nK, limited only by our cooling performance. To achieve
similar resolution using standard methods would require hundreds of
milliseconds of expansion or Bragg beams with tens of Hz frequency
stability. Furthermore, we show evidence of tunneling in a 1D system
when the ``knife-edge'' barrier is spatially thin. This method is
a useful tool for atomic interferometry and for other areas in cold-atom
physics where a robust and precise technique for characterizing the
momentum distribution is crucial. 
\end{abstract}
\maketitle

\section{Introduction\label{sec:Introduction}}

Over the last decades, atomic physics has borrowed techniques and
concepts from optics: atom lasers \cite{Mewes1997,Bloch1999,Hagley1999,Kohl2001,Kohl2005,Billy2007a,Bouyer2002},
atom interferometry \cite{Dickerson2013,Kovachy2015a,McDonald2013,Muntinga2013,M.R.AndrewsC.G.TownsendH.-J.MiesnerD.S.DurfeeD.M.Kurn1997},
and matter-wave lensing \cite{Ammann1997,Chu1986,Marechal1999,Myrskog2000,Morinaga1999},
to mention a few of them. Matter-wave lensing, also known as delta-kick
cooling (DKC), allows for a decrease in the effective temperature
of the atoms by applying a ``lens'' to collimate the atomic cloud.
This technique has been particularly useful for matter-wave interferometry,
where the decrease in temperature translates to longer coherence times.
It is also of interest to other areas where there is a stringent requirement
on the momentum width \cite{Carusotto2001,Jachymski2018}. The temperatures
obtained with delta-kick cooling have been pushed lower and lower
in recent years, achieving temperatures in the sub-nanokelvin regime
\cite{Kovachy2015}, well below standard cooling techniques. This
achievement comes hand in hand with the challenge of measuring such
low temperatures. Standard time-of-flight (TOF) measurements become
inadequate for characterizing the momentum distribution of the atoms,
as the expansion time necessary to determine it precisely increases
to hundreds of milliseconds or even seconds {[}Fig. \ref{fig:setup}b{]}.\\
 \indent In this article, we present an alternative technique to
characterize the momentum distribution. This technique had been envisioned
before in \cite{Billy2007a}. Following an atom-optics approach, this
method relies on performing a knife-edge scan of the momentum distribution
of the atoms with the help of a repulsive potential. A sufficiently
thick barrier transmits only atoms with energies greater than the
potential height, a momentum-space analog of the optical knife-edge.
This method does not depend on the long interrogation times or high
phase stability required by methods such as TOF or Bragg spectroscopy
\cite{Stenger1999}. 

\vfill

\section{Knife-edge technique\label{sec:Knife-edge-technique}}

In optics, a common method to determine a beam radius is to scan a
sharp edge across the beam path and detect the transmitted power.
This technique is called knife-edge measurement. A general expression
for the detected signal is
\begin{equation}
P(x')=\int_{-\infty}^{\infty}I(x)R(x-x')dx,\label{Eq: Convolution}
\end{equation}
where $R(x)$ is the transmission function of the razor blade, and
$I(x)$ is the beam intensity profile. This equation is a convolution
of the transverse profile of the optical beam and the transmission
function of the knife-edge, which for an opaque and sharp razor blade
resembles a step function. Thus the detected signal is the integrated
intensity profile of the optical beam, an error function for a gaussian
beam.

\begin{figure*}[ht]
\includegraphics[width=17cm]{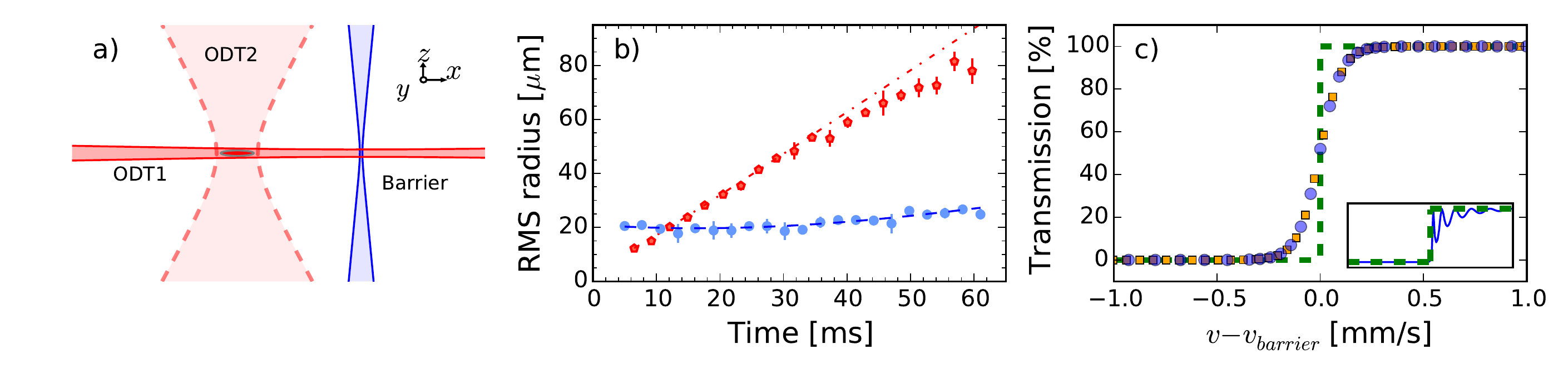}\caption{(a) Experimental setup: atoms are initially trapped in a crossed dipole
trap formed by the ODT1 and ODT2 beams. During the experiment, the
ODT2 beam is turned off, and a magnetic field gradient is used to
push the atoms along the waveguide (ODT1) and towards the barrier.
(b) Time-of-flight measurements of a cloud without cooling with a
temperature of 15 nK (red hexagons), and a cloud after delta-kick
cooling with a temperature of 1.3 nK (blue dots). (c) Calculated transmission
through a thick (3.1 $\mu$m) gaussian barrier for two different barrier
heights: 50 nK (orange squares) and 500 nK (blue dots), and transmission
through a 3.1 $\mu$m square potential (inset). The dashed line is
a step function for comparison with the transmission through a gaussian
barrier (square barrier - inset).}
\label{fig:setup}
\end{figure*}

In the case of an atomic knife-edge measurement, the spatial distribution
of the optical beam is replaced by the velocity distribution of an
atomic wave packet and a gaussian barrier plays the role of the razor
blade. In contrast with a square barrier, the transmission through
a gaussian potential does not exhibit any sharp resonances {[}Fig.
\ref{fig:setup}c{]}, and lacks a closed-form expression. Nevertheless,
for a thick gaussian barrier, the transmission approaches a step function,
thus acting effectively as a high-pass velocity filter. Therefore,
the transmission for a dilute condensate is approximated by

\begin{equation}
T(v)=\frac{1}{2}\left[1+\erf\left(\frac{v-v_{barrier}}{\sqrt{2}\sigma_{at}}\right)\right],\label{Eq: Gaussian transmission}
\end{equation}
where $v$ is the relative velocity of the atoms with respect to the
barrier, $v_{barrier}$ is the velocity corresponding to the barrier
height and $\sigma_{at}$ is the atomic rms velocity width. This expression
is exact for an non-interacting ensemble. Additionally, the measured
rms velocity width $\sigma_{at}$ does not depend on the potential
height {[}Fig. \ref{fig:setup}c{]}, making this technique flexible
as the potential height does not require day to day calibration. 

\section{Experimental system\label{sec:Experimental-system}}

In our experiment, we prepare a\textbf{ }$^{87}$Rb Bose-Einstein
condensate in a 1064 nm crossed dipole trap, with all the atoms in
the $|F=2,m_{F}=2\rangle$ state {[}Fig. \ref{fig:setup}a{]}. One
of the optical dipole trap beams (ODT1) creates a quasi 1D waveguide
for the atoms with trap frequencies $\nu_{r}=217$ Hz and $\nu_{l}=2.7$
Hz, while the other trap beam (ODT2) provides initial confinement
of $\nu_{l_{2}}=61$ Hz along the longitudinal waveguide direction.
After creating a pure BEC, we perform additional forced evaporation
down to $3\times10^{4}$ atoms to lower the interaction energy. This
last evaporation results in a narrower initial momentum distribution
while keeping a high signal-to-noise ratio in the absorption images. 

A thin sheet of light, propagating along the $z$-direction, intersects
the waveguide and produces a repulsive potential {[}Fig. \ref{fig:setup}a{]}.
This potential, described in detail in Ref. \cite{Potnis2017}, is
a 405 nm beam that is focused tightly along the $x$-direction, and
scanned in the $y$-direction using an acousto-optic modulator to
create a flat average potential over a 75 $\mu$m region. This beam
has been characterized outside the experiment, giving a $1/e^{2}$
radius of 1.3 $\mu$m and a Rayleigh range of 8 $\mu$m. By scanning
the barrier along the $z$-direction and colliding the atoms with
it, we locate the waist of the beam. 

\section{Momentum width characterization\label{sec:Momentum-width-characterization}}

We prepare a particular velocity width for the atomic wavepacket through
delta-kick cooling. In short, DKC is the temporal matter-wave analog
of an optical lens: atoms are allowed to expand for a time $T$, then
a harmonic potential with frequency $\omega$ ``kicks'' the atoms
for a duration $\tau$, mimicking a lens with ``focal time'' $f=1/\omega^{2}\tau$.
If $\omega$ and $\tau$ are adjusted such that $f=T$, then the cloud
is collimated, achieving a minimum velocity spread which is reduced
by the ratio of the final cloud to the initial cloud size \cite{Ammann1997,Chu1986,Marechal1999,Myrskog2000,Morinaga1999}.
In our experiment, we realize a two kick sequence that provides finer
control to scan around the best kick duration. This sequence also
yields better performance than a single kick in our setup. The cycle
starts when atoms are released from the crossed dipole trap by turning
the ODT2 beam off and allowed to expand in the waveguide for 12 ms.
This initial expansion time is long enough to convert the interaction
energy into kinetic energy ($t_{exp}>1/\omega_{l2}$). The ODT2 beam
is then flashed for 1 ms, applying an initial kick to the cloud, but
not fully collimating the momentum distribution. The atoms continue
to expand for another 15 ms, and finally, a second kick with half
the power of the first kick is applied for a variable time. The amount
of expansion is limited by the radius of the ODT2 beam (100 $\mu$m):
this expansion time is kept short for the atoms to be within the harmonic
region of the gaussian potential. Ultimately, we found through comparison
with numerical simulations of the Gross-Pitaevskii equation (GPE)
that the cooling efficiency in our setup is not limited by the initial
cloud expansion; a possible explanation is high spatial frequency
perturbations in the delta-kick cooling beam which could lead to forces
comparable to that of the lensing kick \cite{Kovachy2015}.

For the velocity width measurements, the repulsive potential is positioned
off-focus to obtain a barrier width of 3.1 $\mu$m. The thickness
of the potential ensures that tunneling is negligible and the barrier
serves as a sharp momentum filter. Atoms will be transmitted if and
only if their kinetic energy exceeds the barrier height.

After the preparation of the atomic wavepacket, the barrier height
is ramped up to its final value. For the experimental sequence, it
is sufficient to vary either the barrier height or the incident velocity
of the atoms. Given power limitations on the barrier beam, we chose
to scan the incident velocity of the atoms while keeping the barrier
height fixed. A variable strength magnetic field gradient is pulsed
for 0.5 ms along the longitudinal axis of the waveguide to control
the incident velocity of the atoms. For the typical velocities in
the experiment, the ensemble takes 6-8 ms to arrive at the barrier
and 1-3 ms to transverse the barrier. After the interaction is complete,
an absorption image is taken to measure the transmitted and reflected
portions. Finally, using Eq. \ref{Eq: Gaussian transmission}, we
fit this data to extract the barrier height and the velocity width
of the atomic ensemble. 

\subsection{Results \label{subsec:Results}}

We compare our method to the standard TOF measurement. The latter
technique consists of the same DKC sequence, followed by expansion
times up to 60 ms. The maximum expansion time is based on the requirement
that the atoms remain in the waveguide and do not experience any further
lensing from its weak longitudinal harmonic confinement. The fit function
for the rms radius in the TOF measurement is $\Delta x(t)=\sqrt{{{\Delta x_{0}}^{2}+\Delta v^{2}(t-t_{0})^{2}}}$,
where $\Delta x_{0}$ is the minimum atomic rms radius, $\Delta v$
is the atomic rms velocity width and $t_{0}$ is the time at which
the atoms focus to their minimum rms radius. TOF and knife-edge measurements
for a given kick duration are interleaved and randomized to avoid
any possible fluctuations that might affect solely one of the techniques.

\begin{figure}[h]
\includegraphics[width=8.5cm]{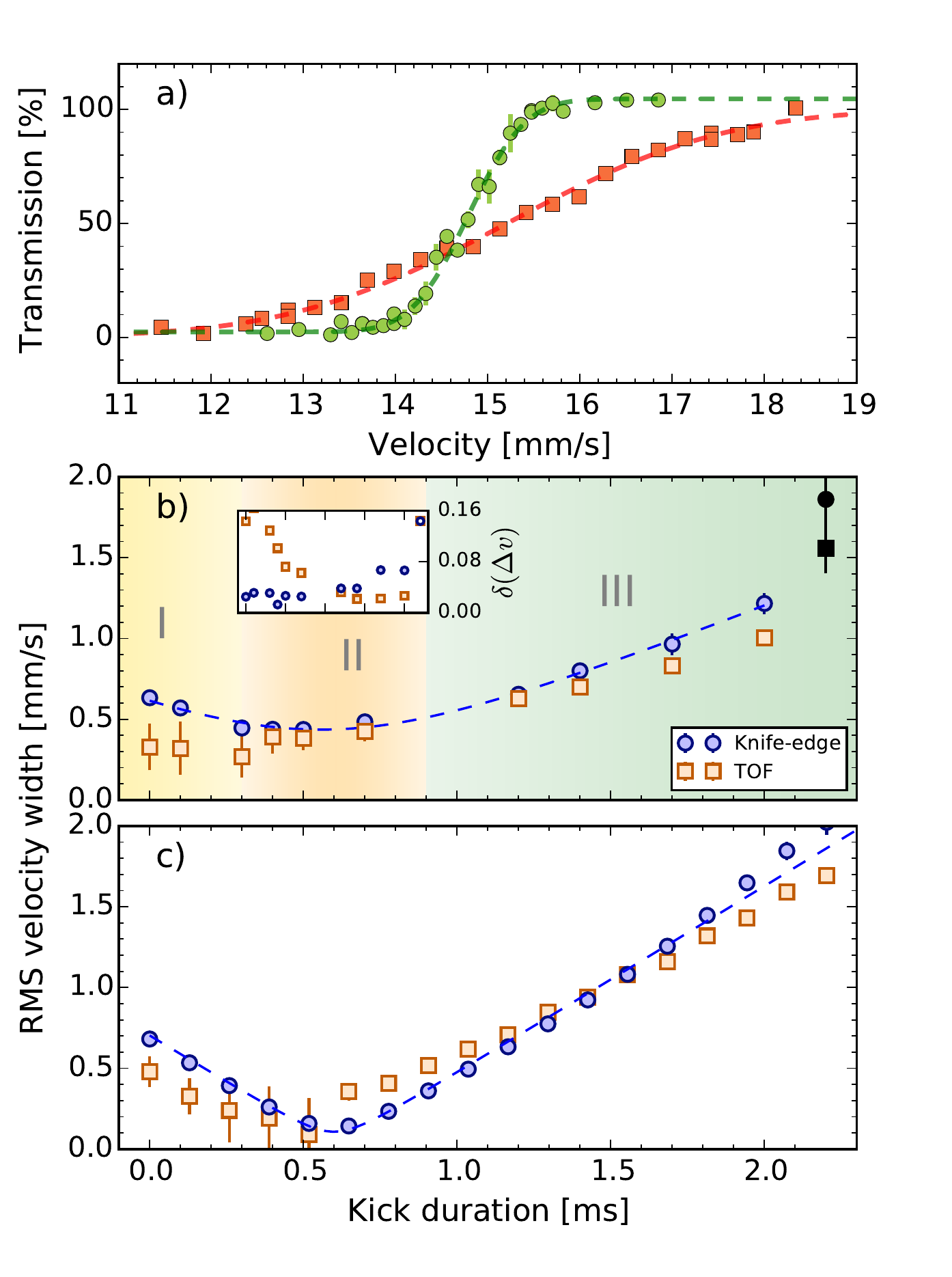}

\caption{Velocity width measurement. (a) Knife edge measurement: transmission
through a thick gaussian barrier for a 0.7 ms kick duration (green
circles) and without cooling (red squares). (b) Different amount of
cooling is performed by adjusting the kick duration of the second
DKC pulse. Blue circles correspond to knife-edge measurements while
red squares are TOF measurements, black solid circle (square) shows
a measurement without the cooling sequence. The dashed blue line is
a quadratic fit, and serves as a guide to the eye. Inset: Uncertainties
for the velocity width measurements in mm/s, horizontal axis same
as Fig. \ref{Fig: Main results}b. (c) 1D Gross-Pitaevskii simulations
for the same velocity width measurements. These simulations follow
the same sequence and analysis as the experimental data. Color coding
is the same as Fig. \ref{Fig: Main results}b.}
\label{Fig: Main results} 
\end{figure}

We identify three distinct regimes when the kick strength is varied
in the cooling sequence: (I) underkicked, (II) close to ideal kick
and (III) overkicked. In the underkicked regime {[}Fig. \ref{Fig: Main results}b,
yellow region{]}, the TOF measurement is not a robust way to obtain
the momentum width as a small amount of noise in the measured cloud
radius can severely affect the ability to estimate the fit parameters
correctly. This can be understood by recognizing that two of the fit
parameters ($\Delta x_{0}$ and $t_{0}$) become highly correlated
in this regime and are difficult to estimate independently, and as
a result, the TOF technique displays large uncertainties. For the
points close to the ideal kick {[}Fig. \ref{Fig: Main results}b,
orange region{]}, the two techniques agree though the TOF measurements
yield uncertainties about three times larger than the knife-edge technique
{[}Fig. \ref{Fig: Main results}b, inset{]}. The TOF error bars are
caused mainly by the constraint of low expansion times, which renders
it difficult to estimate these low-velocity widths precisely. The
resolution of the knife-edge technique is 0.03 mm/s, corresponding
to a temperature resolution of 200 pK. The resolution is limited by
the reproducibility in our cooling sequence, not by the knife-edge
technique. Finally, in the overkicked region {[}Fig. \ref{Fig: Main results}b,
green region{]}, the discrepancy in the measurements comes from the
fast rate of expansion after focusing caused by interactions and the
extra kick due to the weak harmonic confinement along the waveguide.
This additional kick is minimized in the knife-edge technique because
of the short interaction time. These observations are backed up qualitatively
by GPE simulations {[}Fig. \ref{Fig: Main results}c{]}, though a
quantitative comparison would require incorporating the specific imperfections
in our lensing beam, which are not included in these simulations. 

\begin{figure}[h]
\includegraphics[width=8.5cm]{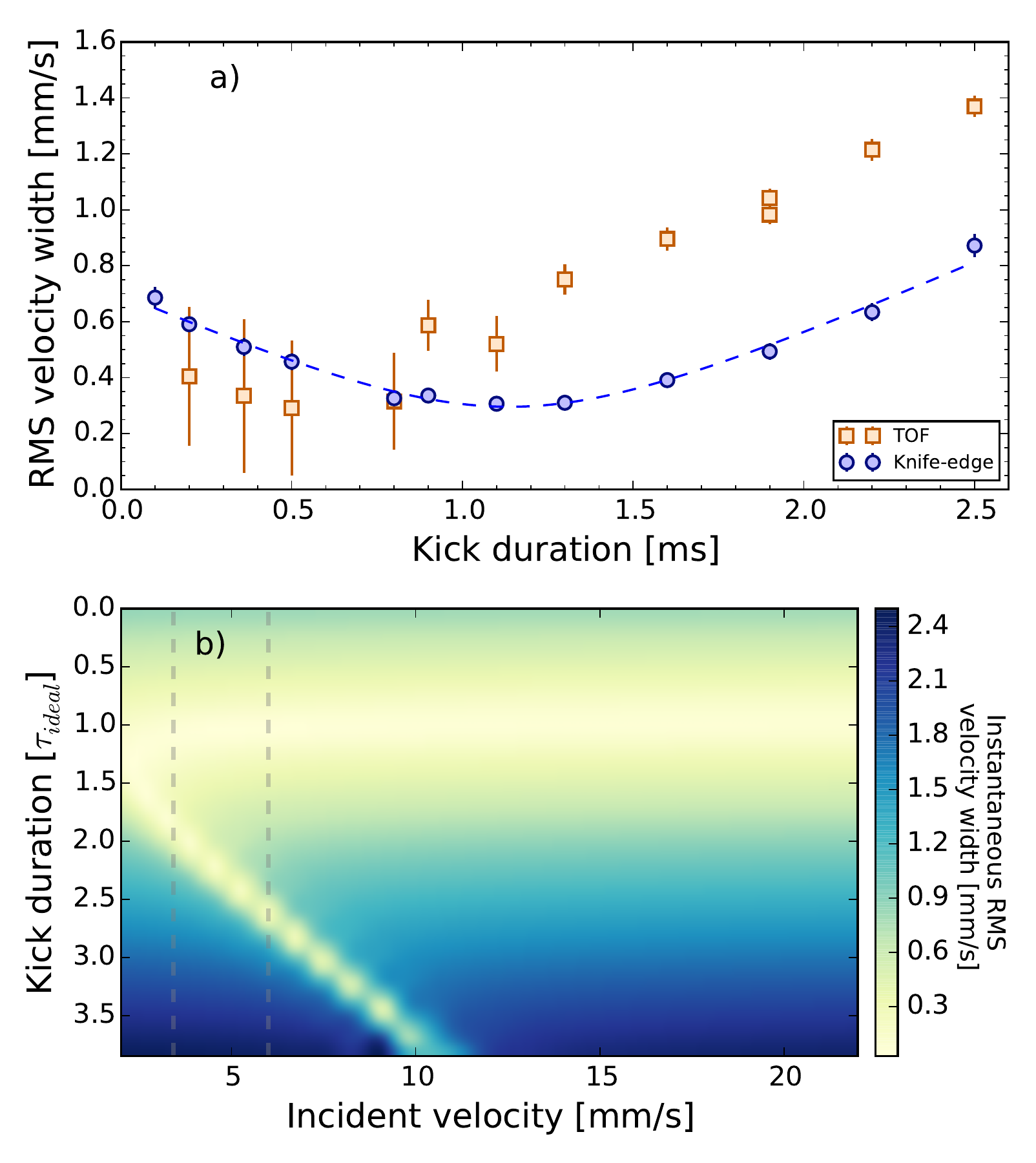}
\caption{Mean field effect on velocity width measurement. (a) Experimental
data. Blue circles correspond to knife-edge measurements taken when
atoms focus at the barrier, which is located 120 $\mu$m away, while
red squares are TOF measurements. Some error bars are smaller than
the symbols. The dashed blue line is a quadratic fit. (b) GPE simulations
of the velocity width at the instant the atoms reach the barrier,
which is determined by the incident velocity (horizontal axis). This
is shown for a range of kick durations (vertical axis). The region
in between the dashed lines is where the points for the knife-edge
measurement in Fig \ref{Fig: Mean field data and sims}a were taken.}

\label{Fig: Mean field data and sims}
\end{figure}

\subsection{Mean-field effect \label{subsec:Mean-field-effect}}

As shown in the previous section, the knife-edge technique is simple
to implement and provides a precise measurement of the atomic momentum
width, but the effect of interactions must be considered as it can
alter the behaviour of a system significantly \cite{Burger1999,Arnold2001,Kashurnikov2001,Potnis2017,Chevy2016}.
In the results from section \ref{subsec:Results}, the effects due
to interactions are negligible, and it is only when the matter-wave
lens focuses the cloud that these effects come into play, due to an
increase in the atomic density.

We have taken additional measurements in the overkicked regime, where
the atom cloud focuses, to explore this effect. The cooling sequence
is similar to the one previously described. The incident velocities
were chosen so that the cloud would collide with the barrier close
to the time at which it is focused to its minimum width. Fig. \ref{Fig: Mean field data and sims}a
shows a series of measurements carried out in the range of incident
velocities indicated by the two gray dashed lines in Fig. \ref{Fig: Mean field data and sims}b.
The knife-edge measurement yields lower velocity widths than the TOF
measurements, and the apparent ideal kick duration is biased towards
longer kick durations due to the points in the region where the kinetic
energy converts to interaction energy. This is confirmed by GPE simulations
{[}Fig. \ref{Fig: Mean field data and sims}b{]} which show that the
instantaneous velocity width decreases considerably in the regions
where the cloud focuses.

\subsection{Comparison with other techniques}

The high precision of the knife-edge technique can be seen in Fig.
\ref{Fig: Main results}. It is only when the velocity width is too
large (sequence without DKC or long kick durations) that the uncertainties
grow due to the scarcity of points at the wings of the transmission
function. The sources of error that could play a role were monitored
in each shot. The fluctuations of the magnetic gradient pulse used
to push the atoms are lower than 1\%, and the power fluctuations on
the barrier beam are kept below 1.5 \% in each knife-edge scan. 

The knife-edge technique shows a clear advantage over the TOF measurement.
To precisely determine the velocity width of the atoms using TOF,
the expansion time has to be much greater than $\Delta x_{o}/\Delta_{v}$,
which in the case of our lowest velocity width corresponds to $\gg$
60 ms. In most cases, the long expansion time restricts this measurement
{[}Fig. \ref{fig:setup}b{]}, with a few notable exceptions, for instance
in experiments where the expansion is conducted in a microgravity
environment \cite{VanZoest2010} or a long vacuum chamber \cite{Zhou2011,Dickerson2013}.

Bragg spectroscopy is another standard tool for obtaining the velocity
spread of an atomic cloud. Atoms moving at a velocity $v$ are diffracted
by two beams with a relative angle $\theta$ and a frequency difference
given by

\begin{equation}
\Delta\omega=\frac{2\hbar k^{2}}{m}+2kv\sin\theta,\label{eq:bragg diffraction}
\end{equation}
where $k$ is the light wavevector, and $m$ is the atomic mass. The
last term corresponds to the Doppler shift, which allows for mapping
of the velocity distribution. Given the obtained atomic velocity width,
the frequency difference to map out the velocity distribution with
a 0.03 mm/s resolution, would correspond to 10's of Hz. The frequency
stability required to diffract the atoms of a particular velocity
class, plus the power stability to maintain the same Rabi frequency,
make this technique demanding for narrow atomic velocity widths.

\section{Tunneling\label{Sec: tunneling}}

Tunneling drives the dynamics in a number of systems in ultracold
atoms: from spatial tunneling in the Bose-Hubbard model \cite{Greiner2002}
to tunneling in phase space in nonlinear systems \cite{Hensinger2001}.
However, a textbook situation where, in a one-dimensional system,
a wavepacket impinges and tunnels through a potential barrier had
not been realized in ultracold atoms before due to the constraints
in the velocity width of the impinging wavepacket, and the exponential
decrease of tunneling with barrier thickness. 

\begin{figure}[h]
\includegraphics[width=8.5cm]{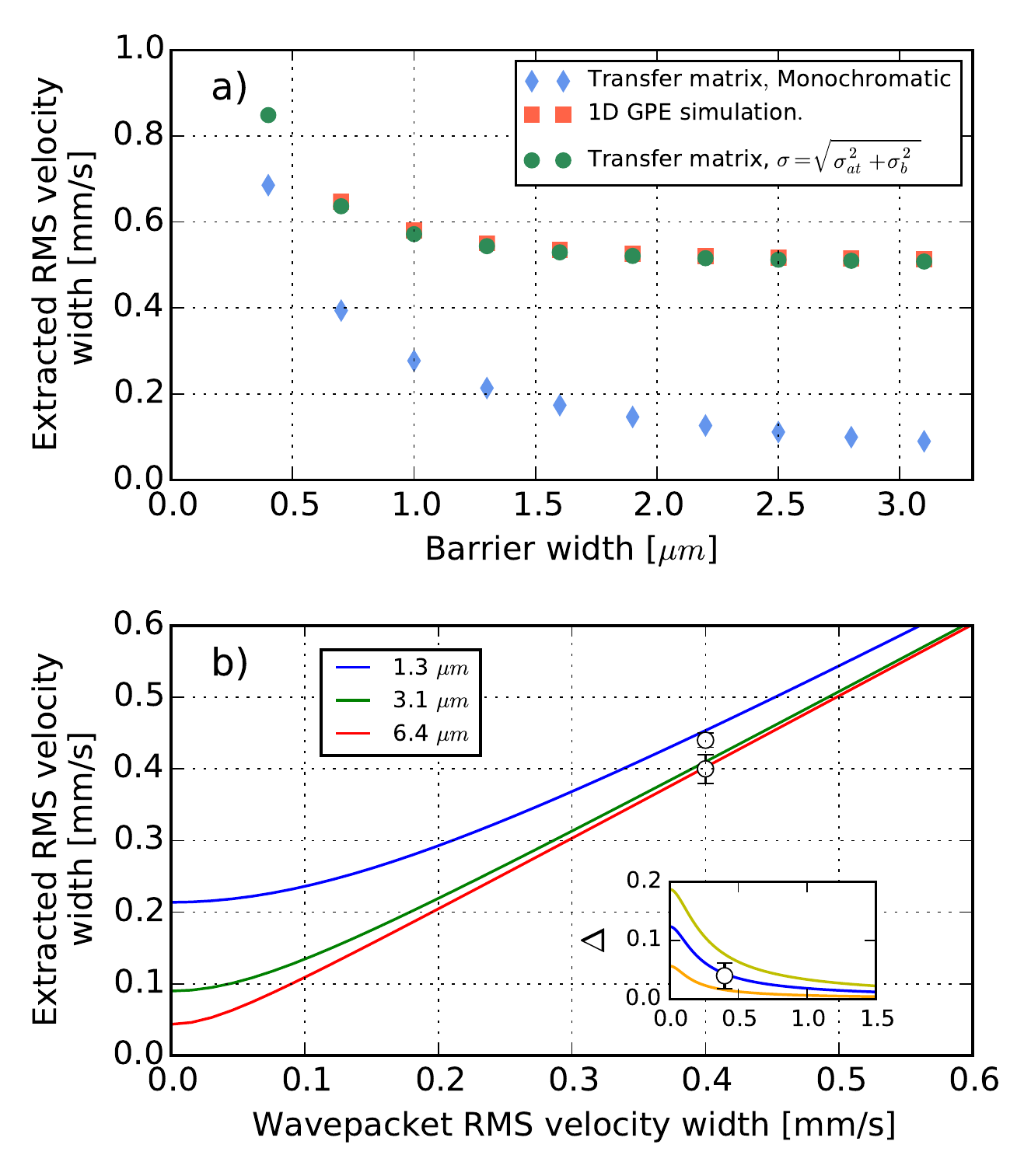}
\caption{(a) Knife-edge extracted rms velocity width as a function of the barrier
width. The simulations were done for a wavepacket with an atomic velocity
width $\sigma_{at}=$ 0.5 mm/s, with the exception of the monochromatic
transfer matrix calculation. (b) Knife-edge extracted rms velocity
against the incident wavepacket velocity width, hollow dots represent
experimental measurements. Inset: Velocity width difference $\Delta$
vs. wavepacket rms velocity width for a 1.0 $\mu$m (yellow), 1.3
$\mu$m (blue) and 1.9 $\mu$m (orange) barrier with a 3.1 $\mu$m
barrier.}
\label{Fig: Tunneling}
\end{figure}

In the previous sections, we discussed the transmission through a
thick barrier and how we can use it to characterize the atomic velocity
width. For a thin barrier, Eq. \ref{Eq: Gaussian transmission} is
no longer valid, as tunneling becomes relevant and modifies the transmission.
The width of the transmission function is given by \textbf{$\sigma=\sqrt{\sigma_{at}^{2}+\sigma_{b}^{2}}$},
where $\sigma_{b}$ depends on the barrier width and accounts for
tunneling in the system. Tunneling causes a thin barrier to behave
as a ``blunt'' knife-edge as it blurs out the cut-off of the atomic
velocity profile. Transfer matrix and 1D GPE simulations of a wavepacket
transmission for different barrier widths are shown in Fig. \ref{Fig: Tunneling}a.
The dependence of $\sigma$ on the barrier width is as expected; tunneling
becomes a more significant correction for thin barriers, while it
vanishes rapidly as the thickness increases, thus recovering the atomic
velocity width.

We have found evidence of tunneling in our system through the dependence
of the observed velocity width on the barrier thickness. Fig. \ref{Fig: Tunneling}b
shows predictions for the measured velocity width as a function of
the wavepacket velocity width. The extracted rms velocity width from
1D GPE simulations agrees with the behaviour expected from the quadrature
sum of the atomic velocity width $\sigma_{at}$ and the contribution
due to tunneling $\sigma_{b}$, where $\sigma_{b}$ for a 1.3 $\mu$m
barrier is about two times greater than that of a 3.1 $\mu$m barrier
{[}Fig \ref{Fig: Tunneling}a - blue diamonds{]}. We measured the
momentum width for the cases of a barrier width of 1.3 $\mu$m and
3.1 $\mu$m {[}Fig. \ref{Fig: Tunneling}b{]}, and found that the
two results differ by two standard deviations. Our measurements agree
with the difference of the momentum width expected from the simulations
{[}Fig. \ref{Fig: Tunneling}b inset{]}, and are thus consistent with
tunneling through a 1.3 $\mu$m barrier.

\section{Conclusions}

We have demonstrated a technique to characterize ultralow velocity
widths corresponding to a resolution in temperature of 200 picokelvin.
We expect this tool to be beneficial for atom interferometry as it
is straightforward to implement, robust in comparison with spectroscopic
techniques, and provides a direct measurement of the velocity width
with high precision on desirable experimental timescales. Additionally,
we observe evidence of tunneling in a quasi 1D system in scattering
configuration. This system should permit novel studies of foundational
questions in quantum mechanics \cite{Steinberg1995,Steinberg1995a}.

\section*{Acknowledgments}

The authors thank Isabelle Racicot and Kent Bonsma-Fisher for critical
reading of the manuscript. R.R. thanks Consejo Nacional de Ciencia
y Tecnología (CONACYT). Computations were performed on the gpc supercomputer
at the SciNet HPC Consortium. SciNet is funded by: the Canada Foundation
for Innovation under the auspices of Compute Canada; the Government
of Ontario; Ontario Research Fund - Research Excellence; and the University
of Toronto. This research was supported by NSERC, CIFAR, Northrop
Grumman Aerospace Systems, and the Fetzer Franklin Fund of the John
E. Fetzer Memorial Trust.

\bibliographystyle{apsrev4-1}
\bibliography{./barrierknife_edge}

\end{document}